\documentclass[
    ,final            
  ]
  {aipproc}
\layoutstyle{8x11double}
\usepackage{clshan-math,clshan-dm-dd}

\begin{document}
\title{Improvement of the Determination of the WIMP Mass \\
       from Direct Dark Matter Detection Data}
\author{Manuel Drees}
      {address={Physikalisches Institut der Universit\"at Bonn, D--53115 Bonn,
      Germany \\ 
                 School of Physics, KIAS, Seoul 130--012, Republic of Korea
      \\ 
                 Bethe Center of Theoretical Physics, Universit\"at Bonn,
      D--53115 Bonn, Germany \\ 
                 E-mail: {\tt drees@th.physik.uni-bonn.de}\\~\\}}
\author{Chung-Lin Shan}
        {address={School of Physics and Astronomy, Seoul National University,
                  Seoul 151--747, Republic of Korea \\
                  E-mail: {\tt cshan@hep1.snu.ac.kr}}}
%
%
\begin{abstract}
  Weakly Interacting Massive Particles (WIMPs) are one of the leading
  candidates for Dark Matter.  We developed a model--independent method for
  determining the WIMP mass by using data (i.e., measured recoil energies) of
  direct detection experiments.  Our method is independent of the as yet
  unknown WIMP density near the Earth, of the form of the WIMP velocity
  distribution, as well as of the WIMP--nucleus cross section.  It requires
  however positive signals from at least two detectors with different target
  nuclei.  At the first phase of this work we found a systematic deviation of
  the reconstructed WIMP mass from the real one for heavy WIMPs.  Now we
  improved this method so that this deviation can be strongly reduced for even
  very high WIMP mass.  The statistical error of the reconstructed mass has
  also been reduced.  In a background--free environment, a WIMP mass of $\sim$
  50 GeV could in principle be determined with an error of $\sim$ 35\% with
  only 2 $\times$ 50 events.
\end{abstract}
\maketitle
%

\section{Introduction}

By now there is strong evidence that more than 80\% of all matter in the
Universe is dark (i.e., interacts at most very weakly with electromagnetic
radiation and ordinary matter). The dominant component of this cosmological
Dark Matter must be due to some yet to be discovered, non--baryonic particles.
Weakly Interacting Massive Particles (WIMPs) $\chi$ with masses roughly
between 10 GeV and a few TeV are one of the leading candidates for Dark Matter
(for reviews, see Refs.~\cite{SUSYDM}).

Currently, the most promising method to detect many different WIMP candidates
is the direct detection of the recoil energy deposited in a low--background
laboratory detector by elastic scattering of ambient WIMPs on the target
nuclei \cite{deta}.  The differential rate for elastic WIMP--nucleus
scattering is given by \cite{SUSYDM}:
\beq \label{eqn:dRdQ}
   \dRdQ
 = \calA \FQ \int_{v_{\rm min}}^{\infty} \bfrac{f_1(v)}{v} dv\, .
\eeq
Here $R$ is the direct detection event rate, i.e., the number of events per
unit time and unit mass of detector material, $Q$ is the energy deposited in
the detector, $F(Q)$ is the elastic nuclear form factor, $f_1(v)$ is the
one--dimensional velocity distribution function of the WIMPs impinging on the
detector, $v$ is the absolute value of the WIMP velocity in the laboratory
frame.  The constant coefficient $\calA$ is defined as
\beq \label{eqn:calA}
        \calA
 \equiv \frac{\rho_0 \sigma_0}{2 \mchi m_{\rm r,N}^2}\, ,
\eeq
where $\rho_0$ is the WIMP density near the Earth and $\sigma_0$ is the total
cross section ignoring the form factor suppression.  The reduced mass $m_{\rm
  r,N}$ is defined by
\beq \label{eqn:mrN}
        m_{\rm r,N}
 \equiv \frac{\mchi \mN}{\mchi+\mN}\, ,
\eeq
where $\mchi$ is the WIMP mass and $\mN$ that of the target nucleus.  Finally,
$\vmin = \alpha \sqrt{Q}$ is the minimal incoming velocity of incident WIMPs
that can deposit the energy $Q$ in the detector and
\beq \label{eqn:alpha}
        \alpha
 \equiv \sfrac{\mN}{2 m_{\rm r,N}^2}\, .
\eeq

In our earlier work \cite{DMDDf1v}, we developed methods for reconstructing
the one--dimensional velocity distribution, $f_1(v)$, and for estimating its
moments from the recoil spectrum as well as from measured recoil energies
directly in direct detection experiments:
\beqn \label{eqn:moments}
           \expv{v^n}
 \eqnequiv \int_{v_{\rm min}(\Qthre)}^\infty v^n f_1(v) \~ dv
           \non\\
 \=        \alpha^n \!\!
           \bfrac{2 \Qthre^{(n+1)/2} \rthre/\FQthre+(n+1) I_n}
                 {2 \Qthre^{1    /2} \rthre/\FQthre+      I_0}\!\!.
\eeqn
Here $\rthre \equiv (dR/dQ)_{Q = \Qthre}$ is an estimated value of the
scattering spectrum at the threshold energy, $\Qthre$, and $I_n$ can be
estimated by
\beq \label{eqn:In_sum}
   I_n
 = \sum_a \frac{Q_a^{(n-1)/2}}{F^2(Q_a)}\, ,
\eeq
where the sum runs over all events in the data set. Note that
Eq.(\ref{eqn:moments}) can be used without prior knowledge of the local WIMP
density, $\rho_0$, of the velocity distribution function of incident WIMPs,
$f_1(v)$, as well as of the WIMP--nucleus cross section, $\sigma_0$.

\section{Determining the WIMP mass}

By requiring that the values of a given moment of $f_1(v)$ estimated by
Eq.(\ref{eqn:moments}) from two detectors with different target nuclei, $X$
and $Y$, agree, we found a simple expression for determining the WIMP mass
\cite{DMDDmchi-SUSY07}:
\beq \label{eqn:mchi_Rn} 
   \mchi
 = \frac{\sqrt{\mX \mY}-\mX ({\cal R}_{n,X}/{\cal R}_{n,Y})}
        {{\cal R}_{n,X}/{\cal R}_{n,Y}-\sqrt{\mX/\mY}}\, ,
\eeq
with
\beqn \label{eqn:Rn_thre}
    {\cal R}_{n,X}
 \= \!\!
    \bfrac{2 \QthreX^{(n+1)/2} \rthreX/\FQthreX+(n+1) \InX}
          {2 \QthreX^{   1 /2} \rthreX/\FQthreX+      \IzX}^{1/n}\!\!\!\!\! .
    \non\\
\eeqn
Here $n \ne 0$, $m_{(X,Y)}$ and $F_{(X,Y)}(Q)$ are the masses and the form
factors of the nuclei $X$ and $Y$, respectively, and $r_{{\rm thre},(X,Y)}$
refer to the counting rates at the threshold energies of two detectors.  Note
that the form factors in the estimates of $\InX$ and $\InY$ using
Eq.(\ref{eqn:In_sum}) are different.

Additionally, for the spin--independent (SI) scattering of a supersymmetric
neutralino, which is the perhaps best motivated WIMP candidate \cite{SUSYDM},
and for all WIMPs which interact primarily through Higgs exchange, the SI
scattering cross section is approximately the same for both protons p and
neutrons n. Writing the ``pointlike'' cross section $\sigma_0$ of
Eq.(\ref{eqn:calA}) as
\beq \label{eqn:sigma0SI}
   \sigma_0
 = \afrac{4}{\pi} m_{\rm r,N}^2 A^2 |f_{\rm p}|^2\, ,
\eeq
where $f_{\rm p}$ is the effective $\chi \chi {\rm p p}$ four--point coupling,
$A$ is the atomic number of the target nucleus, one finds
\beq \label{eqn:rho_fp2}
   \rho_0 |f_{\rm p}|^2
 = \frac{\pi}{4 \sqrt{2}} \afrac{\mchi+\mN}{\calE A^2 \sqrt{\mN}}
   \bbrac{\frac{2 \Qthre^{1/2} \rthre}{\FQthre}+I_0}\, .
\eeq
The factor $\calE$ appearing in the denominator is the exposure of the
experiment, which is dimensionless in natural units; it relates the actual
counting rate to the normalized rate (\ref{eqn:dRdQ}). Since the unknown
factor $\rho_0 |f_{\rm p}|^2$ on the left--hand side here is identical for
different targets, Eq.(\ref{eqn:rho_fp2}) leads to another expression for
determining $\mchi$:
\beq \label{eqn:mchi_Rsigma}
   \mchi
 = \frac{\abrac{\mX/\mY}^{5/2} \mY-\mX (\calR_{\sigma,X}/\calR_{\sigma,Y})}
        {\calR_{\sigma,X}/\calR_{\sigma,Y}-\abrac{\mX/\mY}^{5/2}}\, .
\eeq
Here we have assumed $m_{(X,Y)} \propto A_{(X,Y)}$, and introduced the
quantity
\beq \label{eqn:Rsigma_thre}
\calR_{\sigma,X}
 = \frac{1}{\calE_X} \bbrac{\frac{2 \QthreX^{1/2} \rthreX}{\FQthreX}+\IzX}\, .
\eeq

In order to yield the best--fit value of $\mchi$ as well as its statistical
error, we introduced a $\chi^2$ function:
\beq \label{eqn:chi2}
\chi^2 = \sum_{i,j} \abrac{f_{i,X}-f_{i,Y}} {\cal C}^{-1}_{ij}
\abrac{f_{j,X}-f_{j,Y}}\, . 
\eeq
It combines the estimators for different $n$ in Eq.(\ref{eqn:mchi_Rn}) with
each other, and with the estimator in Eq.(\ref{eqn:mchi_Rsigma}). In
Eq.(\ref{eqn:chi2}) we defined
\cheqna
\beqn \label{eqn:fiXa}
    f_{i,X}
  \= \afrac{\alpha_X {\cal R}_{i,X}}{300~{\rm km/s}}^{i}\, ,
\eeqn
for $i = -1,~1,~2,~\dots,~n_{\rm max}$; and
\cheqnb
\beqn \label{eqn:fiXb}
    f_{n_{\rm max}+1,X}
 \= \frac{A_X^2}{\calR_{\sigma,X}}
    \afrac{\sqrt{\mX}}{\mchi+\mX}\, ;
\eeqn
\cheqn
we analogously defined also $n_{\rm max} + 2$ functions $f_{i,Y}$.  Here
$n_{\rm max}$ determines the highest moment of $f_1(v)$ that is included in
the fit.  The $f_i$ are normalized such that they are dimensionless and very
roughly of order unity; this alleviates numerical problems associated with the
inversion of their covariance matrix.  The first $n_ {\rm max}+1$ functions
$f_i$ are basically our estimators of the moments in Eq.(\ref{eqn:moments}).
The last function is essentially the right--hand side in
Eq.(\ref{eqn:rho_fp2}).  It is important to note that $\mchi$ in
Eqs.(\ref{eqn:fiXa}) and (\ref{eqn:fiXb}) is now a fit parameter, not the true
(input) value of the WIMP mass.  Recall also that our estimator
(\ref{eqn:In_sum}) for $I_n$ appearing in Eqs.(\ref{eqn:fiXa}) and
(\ref{eqn:fiXb}) is independent of $\mchi$.  Hence the first $n_{\rm max}+1$
fit functions depend on $\mchi$ only through the overall factor $\alpha^i$.

Moreover, $\cal C$ here is the total covariance matrix.  Since the $X$ and $Y$
quantities are statistically completely independent, $\cal C$ can be written
as a sum of two terms:
\beq \label{eqn:Cij}
   {\cal C}_{ij}
 = {\rm cov}\abrac{f_{i,X},f_{j,X}}+{\rm cov}\abrac{f_{i,Y},f_{j,Y}}\, .
\eeq
The entries of this matrix involving only the moments of the WIMP velocity
distribution can be read off Eq.(82) of Ref.~\cite{DMDDf1v}, with an obvious
modification due to the normalization factor in Eq.(\ref{eqn:fiXa}).  Since
the last $f_i$ defined in Eq.(\ref{eqn:fiXb}) can be computed from the same
basic quantities, i.e., the counting rates at $\Qthre$ and the integrals
$I_0$, the entries of the covariance matrix involving this last fit function
can also be computed straightforwardly.

\section{Matching the cut--off energies}

The basic requirement of our method for determining $\mchi$ given in
Eq.(\ref{eqn:mchi_Rn}) is that, from two experiments with different target
nuclei, the values of a given moment of the WIMP velocity distribution
estimated by Eq.(\ref{eqn:moments}) should agree. This means that the upper
cuts on $f_1(v)$ in two data sets should be (approximately)
equal.\footnote{Here we assume the threshold energies to be negligibly small.}
Since $v_{\rm cut} = \alpha \sqrt{Q_{\rm max}}$, it requires that
\beq \label{eqn:match} 
Q_{\rm max,Y} = \afrac{\alpha_X}{\alpha_Y}^2 Q_{\rm max,X}\,.  
\eeq 
Note that $\alpha$ defined in Eq.(\ref{eqn:alpha}) is a function of the true
WIMP mass.  Thus this relation for matching optimal cut--off energies can be
used only if $\mchi$ is already known.  One possibility to overcome this
problem is to fix the cut--off energy of the experiment with the heavier
target, minimize the $\chi^2(m_{\chi,{\rm rec}})$ function defined in
Eq.(\ref{eqn:chi2}), and estimate the cut--off energy for the lighter nucleus
by Eq.(\ref{eqn:match}) algorithmically.

In Figs.~1 we show our numerical results for the reconstructed WIMP mass based
on Monte Carlo simulations.  $\rmXA{Ge}{76}$ and $\rmXA{Si}{28}$ have been
chosen as two targets.  The dotted (green) lines show results for $Q_{\rm max,
  Ge} = Q_{\rm max, Si} = 100$ keV with a systematic deviation discussed in
Ref.~\cite{DMDDmchi-SUSY07}, whereas the solid (black) lines have been
obtained by using Eq.(\ref{eqn:match}) with $Q_{\rm max, Ge} = 100$ keV and
the true (input) WIMP mass $m_{\chi,{\rm in}}$.  The dashed (red) lines are
for the case that $Q_{\rm max, Ge} = 100$ keV, whereas $Q_{\rm max, Si}$ has
been chosen such that $\chi^2(m_{\chi,{\rm rec}})$ is minimal.  As shown here,
with only 50 events (upper frame) from one experiment, our algorithmic process
seems already to work pretty well for WIMP masses up to $\sim 500$ GeV.
\begin{figure}[t!]
\vspace*{-1cm}
\rotatebox{-90}{\includegraphics[width=0.75\columnwidth]{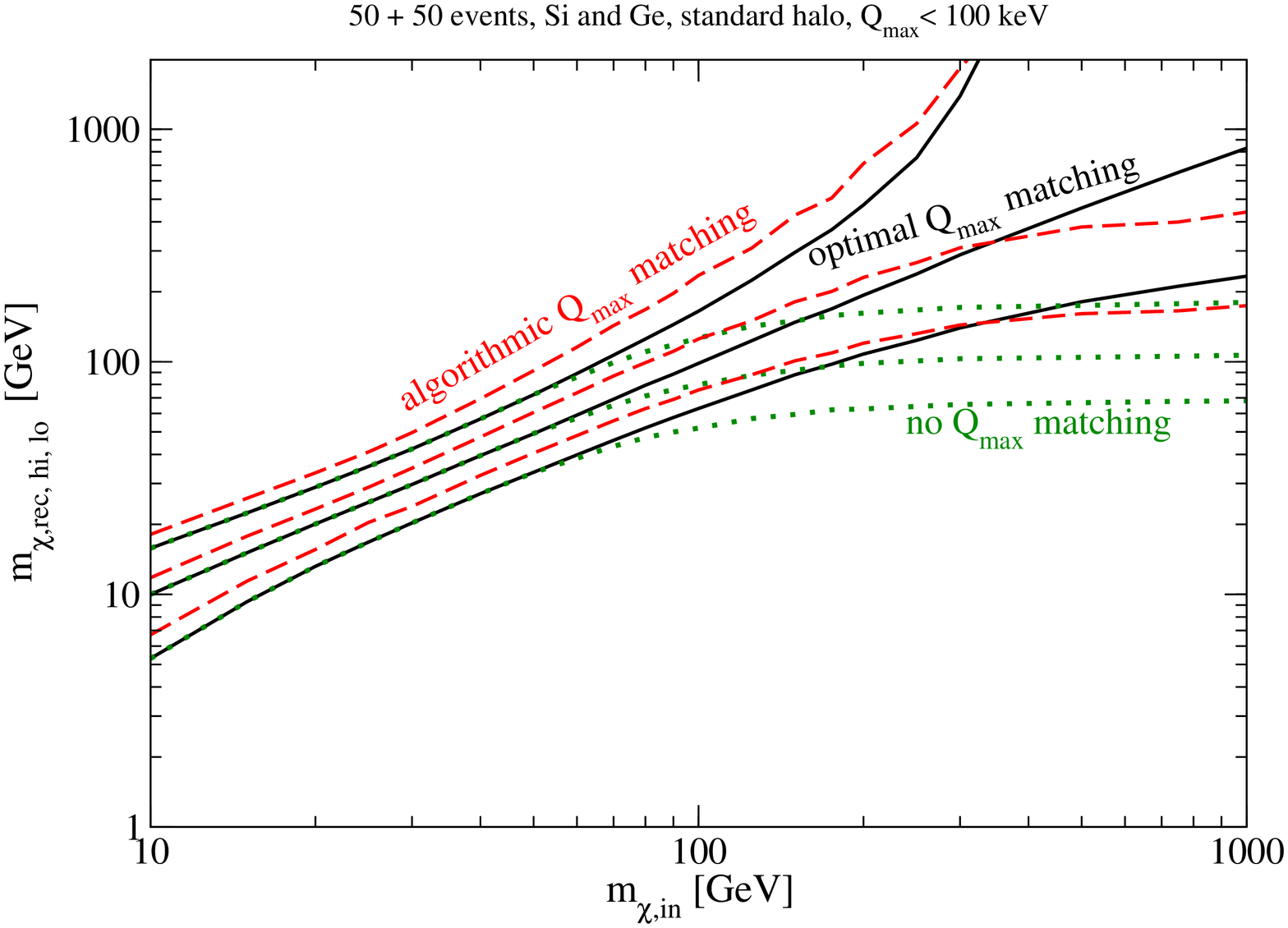}
                \includegraphics[width=0.75\columnwidth]{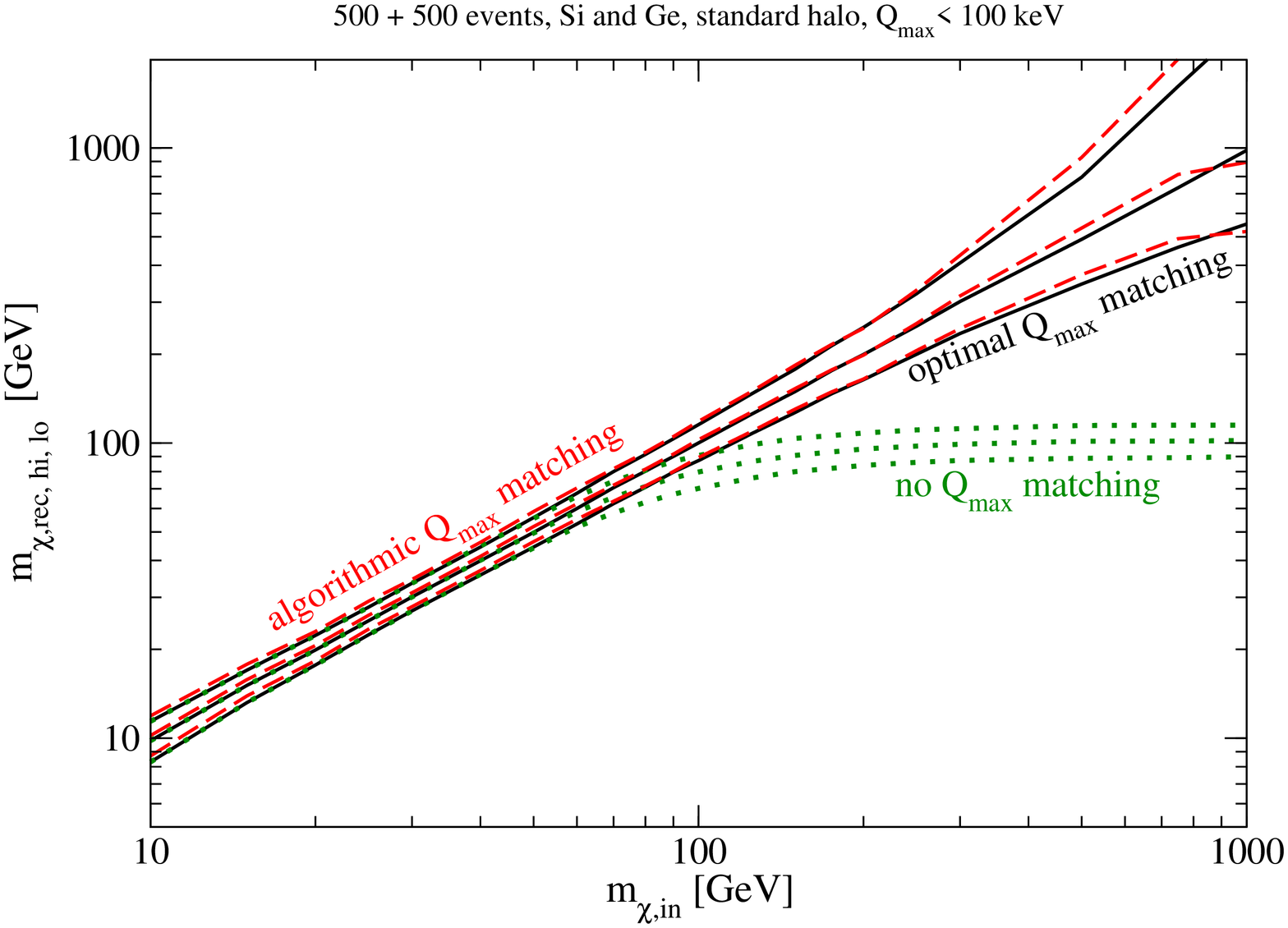}}
\vspace*{-1cm}
\caption{
  Results for the reconstructed WIMP mass as well as its error interval based
  on the combined fit, using Eq.(\ref{eqn:chi2}) with $n_{\rm max} = 2$.  We
  assume that the scattering cross section is dominated by spin--independent
  interactions, The theoretical predicted recoil spectrum for the shifted
  Maxwellian velocity distribution function \cite{SUSYDM}, \cite{DMDDf1v} with
  Woods-Saxon elastic form factor \cite{Engel91}, \cite{SUSYDM} ($v_0 = 220$
  km/s, $v_e = 231$ km/s) have been used.  We simulated with 50 (upper) and
  500 (lower) events on average from each experiment before cuts.  }
\end{figure}

\section{Summary}

In this paper we described the basic ideas of our method for determining the
WIMP mass and gave the main formulae.  The algorithmic process for correcting
the systematic deviation of the reconstructed WIMP mass has also been
discussed.
More details and discussions about determining $\mchi$ can be found in
Ref.~\cite{DMDDmchi}. 
%
%
%
%
%
%
\begin{theacknowledgments}
  This work was partially supported by the Marie Curie Training Research
  Network ``UniverseNet'' under contract no.~MRTN-CT-2006-035863 as well as by
  the European Network of Theoretical Astroparticle Physics ENTApP ILIAS/N6
  under contract no.~RII3-CT-2004-506222.
\end{theacknowledgments}
%
%
%
\bibliographystyle{aipproc}   
%
%
\bibliography{sample}
%
%
\IfFileExists{\jobname.bbl}{}
 {\typeout{}
  \typeout{******************************************}
  \typeout{** Please run "bibtex \jobname" to optain}
  \typeout{** the bibliography and then re-run LaTeX}
  \typeout{** twice to fix the references!}
  \typeout{******************************************}
  \typeout{}
 }
%
%
%

%
\end{document}